\documentclass[12pt,preprint]{emulateapj}

\shorttitle{Contact Binary Trojan Asteroids}
\shortauthors{Mann, Jewitt \& Lacerda}

\received{2006 October 6}
\slugcomment{{\sc Accepted to AJ:} 2007 May 22}
\begin{document}

\title{Fraction of Contact Binary Trojan Asteroids}
\author{Rita K. Mann, David Jewitt \& Pedro Lacerda}

\affil{Institute for Astronomy, University of Hawaii, 2680 Woodlawn Drive, Honolulu, HI 96822}
\email{rmann@ifa.hawaii.edu, jewitt@ifa.hawaii.edu, pedro@ifa.hawaii.edu}

\begin{abstract}
We present the results of an optical lightcurve survey of 114 Jovian 
Trojan asteroids conducted to determine the fraction of contact 
binaries.  Sparse-sampling was used to assess the photometric range
of the asteroids and those showing the largest ranges were targeted
for detailed follow-up observations.  This survey led to the discovery 
of two Trojan asteroids, (17365) and (29314) displaying large lightcurve 
ranges ($\sim$ 1 magnitude) and long rotation periods ($<$ 2 rotations 
per day) consistent with a contact binary nature.  The optical 
lightcurves of both asteroids are well matched by Roche binary 
equilibrium models.  Using these binary models, we find low densities 
of $\sim$ 600 kg m$^{-3}$ and 800 kg m$^{-3}$, suggestive of porous 
interiors.  The fraction of contact binaries is estimated to be 
between 6\% and 10\%, comparable to the fraction in the Kuiper Belt.  
The total binary fraction in the Trojan clouds (including both wide 
and close pairs) must be higher.
\end{abstract}

\keywords{minor planets --- asteroids --- solar system: general --- surveys}

\section{Introduction}\label{sec: intro}
The existence and importance of binary asteroids in small-body 
populations has only been realized in the last decade, after the first 
unambiguous detection of a satellite around main-belt asteroid 243 
Ida by the Galileo spacecraft \citep{belton,chapman}.  It is now 
evident that binaries exist in the main-belt asteroids, the near-earth 
asteroids and in the Kuiper Belt (see review by \citet{richardson06} 
and references therein).  Apart from spacecraft flybys 
(and the rare case of measuring
gravitational perturbations of planets by very large asteroids), studying 
the orbital dynamics of binary systems provides the only method 
available for calculating mass and density.  Density measurements 
are important as probes of internal structure, enabling constraints 
to be placed on the porosity and composition.

The Jovian Trojan asteroids are trapped in a 1:1 mean
motion resonance with Jupiter.  They form two large
clouds around the stable (L4, L5) Lagrangian points 60\degr\ ahead of and behind
the giant planet.  It has been estimated that $\sim$ 10$^5$ Trojan asteroids
with diameters larger than 1-km exist \citep{jewitt00,yoshida}, comparable in 
number to the Main Belt population 
($6.7 \times 10^5$ asteroids, \citet{ivezic}), making it clear 
that they comprise an important reservoir of information.
The Trojan asteroids of Jupiter have yet to be searched systematically for
the presence of binaries.  Despite this fact, two Trojan binaries
have already been identified:
617 Patroclus, a resolved wide binary discovered by
Merline et al. (2001),
while 624 Hektor has a distinctive lightcurve that
indicates it is a close or contact binary (\cite{cook71}, \cite{hartmann88}) and a widely
separated satellite has recently been imaged \citep{marchis_iauc}.
The Trojans are intriguing because they show 
larger photometric ranges when compared with 
main-belt asteroids \citep{hartmann88}, particularly those with 
diameters larger than 90-km \citep{binzel}.  
Large lightcurve amplitudes suggest
elongated shapes or binarity.

While it is not clear whether the Trojans formed at their current 
location alongside Jupiter or were trapped after forming at larger
heliocentric distances \citep{morbidelli}, it is believed
that these bodies are primordial.
Understanding their composition and internal structure
is therefore of great interest, making density determination vital.
The density of Trojan 617 Patroclus has been estimated as
 $\rho = 800^{+200}_{-100}$ kg m$^{-3}$ based on the measured orbital
 period and size, and on diameter determinations made from infrared
 data \citep{marchis}.  This low density contrasts with a comparatively high
 estimate for 624 Hektor, namely $\rho$ = 2480$^{+290}_{-80}$ kg m$^{-3}$,
 determined from the lightcurve and a Roche binary model \citep{lacerda}.

Close or contact binaries are composed of two asteroids 
in a tight orbit around each other.  The Trojan contact binary 
fraction is potentially important in distinguishing between various 
formation theories.  For example, one model of binary formation by
dynamical friction predicts that close binaries should be common 
\citep{goldreich} while another based on 3-body interactions asserts that
they should be rare \citep{weidenschilling02}.  The nature of the Trojan
binaries can also reveal clues about their formation.  It is known that 
different mechanisms formed binaries in the Main Belt and the Kuiper Belt
because of the distinct types of binaries found in both populations.
It is suspected that gravitational processes predominantly form
Kuiper Belt binaries, the known examples of which have components of comparable
mass and large separations \citep{weidenschilling02,goldreich,funato,astakhov}.
Sub-catastrophic impacts followed by
gravitational interaction with the debris formed are
the leading way to form tight binary systems with unequal mass
components that make up the larger main-belt binary population
\citep{weidenschilling89,richardson06}.
A comparative study of the binaries in the Trojan clouds, the Main Belt and
the Kuiper Belt might illuminate the different roles played
by formation conditions in these populations.

Motivated by the lack of studies about Trojan binaries, the aim of this paper
is to investigate the fraction of close or contact binary systems
among the Jovian Trojan population.  Contact binaries 
are specifically targeted for the ease with which they can be identified 
using optical lightcurve information.  
Here, we present a technique called sparse sampling, which 
we used to conduct a lightcurve survey of 114 Jovian Trojan 
asteroids.  The results of this survey, the discovery of two
suspected contact binary asteroids and a discussion of the 
binary fraction in the Jovian Trojan population will follow.

\section{Observations}

\subsection{Sparse Sampling}
The maximum photometric range that can be exhibited by a rotationally elongated, strengthless 
body is 0.9 mag \citep{leone}.  Ranges larger than 0.9 mag. are strongly suggestive of a
contact binary nature, in which mutual
gravitational deformation of the components can drive the range
up to $\sim$ 1.2 magnitudes \citep{weidenschilling80, leone}.  In principle, structurally strong bodies can maintain 
any shape and show an arbitrarily large photometric range.  However, most main-belt asteroids larger than $\sim$ 150-m in 
diameter show little sign of possessing internal strength sufficient to resist 
gravity and/or rotational deformation \citep{pravec, holsapple04} and we expect that the Trojan asteroids are 
similarly structurally weak.  In what follows, we assume that objects with photometric range $>$0.9 mag.
are candidate contact binaries.

\begin{figure}
\epsscale{1.0}
\plotone{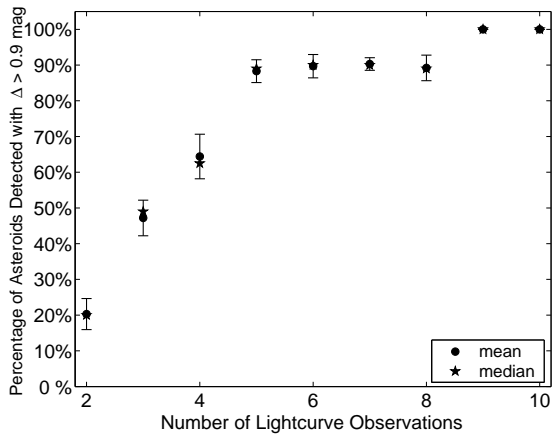}
\caption{Percentage of asteroids detected with photometric
ranges greater than 0.9 magnitudes versus number of lightcurve
observations.  Monte Carlo simulations were conducted
on a sample of asteroids with a photometric
range of 1.2 magnitudes and single-peaked lightcurve
periods between 3 and 10 hours to determine sparse
sampling efficiency.}
\end{figure}

To examine the efficiency of sparse lightcurve sampling, we conducted
a series of Monte Carlo tests.  The tests were applied to asteroids 
with a photometric range of 1.2 magnitudes and double-peaked lightcurve 
periods uniformly distributed between 6 and 20 hours.  The lightcurves 
were uniformly sampled by N=1,2...10 observations over one night.  
Asteroids for which the sparse-sampling technique detected photometric 
ranges between 0.9 and 1.2 magnitudes were picked out as successful 
candidates.  Monte Carlo simulations suggest that between 85\% and 
92\% of asteroids with photometric ranges of 1.2 magnitudes would be 
identified as contact binary candidates from just five measurements 
of brightness per night (see Figure 1).  (The efficiency of detecting 
brightness variations larger than 0.9 magnitudes ranged from 
$\sim$ 71\% for asteroids with actual peak-to-peak lightcurve
amplitudes of 1.0 magnitudes to $\sim$ 81\% of asteroids with
peak-to-peak amplitudes of 1.1 magnitudes.)  The simulations indicate that 
the accuracy with which contact binary candidates are identified varies 
little when sampling between five and eight lightcurve points per 
asteroid (see Figure 1).  
The advantage of sparse sampling is clear: estimates
of photometric range for a large number of asteroids can be made
rapidly, significantly reducing observing time.  Asteroids exhibiting 
large photometric ranges in the sparse sampling study are
subsequently targeted for detailed follow-up observations with 
dense coverage in rotational phase space.  

To further test the sparse sampling technique, we 
observed 2674 Pandarus and 944 Hidalgo, two asteroids known to 
show large photometric variations. 
From published lightcurves, 2674 Pandarus is known to have a 
photometric range of 0.49 magnitudes \citep{hartmann88}.
Using the sparse sampling technique, with the same sampling 
as for all other asteroids in the study (and without prior knowledge of the
rotational phase), we measured a lightcurve
amplitude of 0.50 $\pm$ 0.01 magnitudes for Pandarus.  
Hidalgo has shown a maximum photometric 
variation of 0.60 magnitudes \citep{harris}, whereas
sparse sampling measured the brightness range to be
0.58 $\pm$ 0.02 magnitudes (see Figures 2 and 3).
The agreement  results show that the
photometric range can be usefully estimated with only five
measurements of asteroid brightness.  

Having gained confidence in the technique through simulations and
observational tests, we applied sparse sampling to the Trojan asteroids.
Taking five short exposures, while cycling through the asteroids,
we were able to obtain limited sampling of 114 asteroid
lightcurves in nine good weather nights of observing.

\begin{figure}
\epsscale{1.0}
\plotone{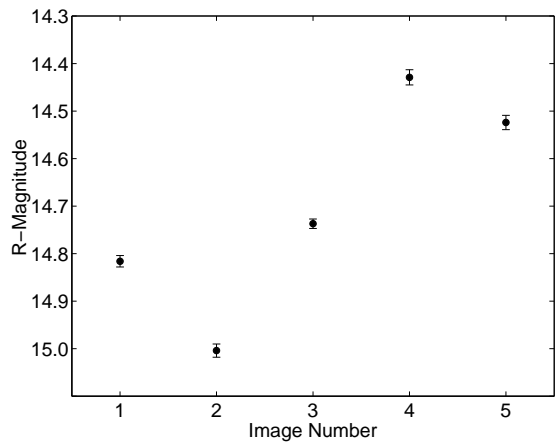}
\caption{Sparse-sampled R-band photometry of 944 Hidalgo.  The photometric 
range estimated from five observations is 0.58 $\pm$ 0.02 
magnitudes, consistent with previous measurements of
0.60 magnitudes from Harris et al. (2006).}
\end{figure}

\begin{figure}
\epsscale{1.0}
\plotone{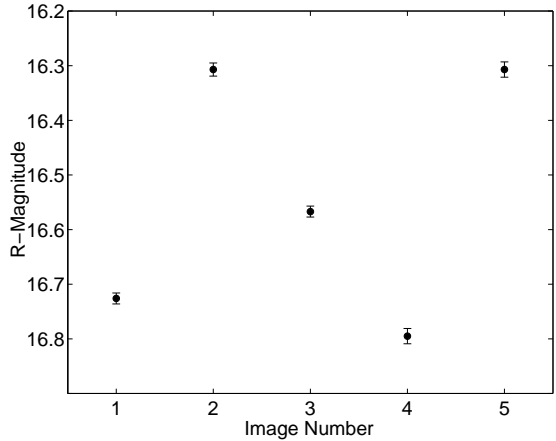}
\caption{Sparse-sampled R-band photometry of 2674 Pandarus. 
The photometric range estimated from five observations is
0.50 $\pm$ 0.01 magnitudes, consistent with previous
measurements of 0.49 magnitudes (Hartmann et al. 1988).}
\end{figure}

\subsection{Data Acquisition and Reduction}
We obtained sparsely sampled optical lightcurve data for the
Jovian Trojan asteroids using both the University of Hawaii 2.2-m
telescope on Mauna Kea and the Lulin One-meter Telescope (LOT) in
Taiwan.  We used a 2048 x 2048 pixel Tektronix charge-coupled 
device (CCD) on the 2.2-m telescope.  This detector has a 
0.219 arcseconds per pixel image scale and a field of view of 
7.5 square arcminutes.  The CCD
on LOT (VersArray:1300B) has 1340 x 1300 pixels with 0.516
arcseconds per pixel scale, and a field of view of 11.5 x 11.2
arcminutes.  All images
were taken in the R band with exposure times scaled to the
brightnesses of the asteroids.  On LOT, the exposure times ranged
from 30 seconds for objects brighter than 15th magnitude, up to
120 seconds for 19th magnitude Trojans.  At the 2.2-m telescope,
the exposure times ranged from 10 seconds for objects brighter
than 17th magnitude, to 150 seconds for 20th magnitude asteroids.
See Table 1 for a description of the observations.  
\LongTables
\begin{deluxetable*}{lccccr}
\tablecolumns{6}
\tablewidth{0pc}
\tabletypesize{\scriptsize}
\tablecaption{Journal of Observations}
\tablehead{
\colhead{UT Date} & \colhead{Telescope} & \colhead{Seeing (\arcsec)} & \colhead{Project\tablenotemark{a}} & \colhead{Full/Half Night} & \colhead{Comments}}
\startdata
2005 March 07 & LOT 1-m & 2.0 & Sparse & Full & Scattered Cirrus \\
2005 March 09 & LOT 1-m & 2.2 & Sparse & Full & Windy\\
2005 March 11 & LOT 1-m & 2.0 & Sparse & Half & Cloudy\\
2005 March 13 & LOT 1-m & 1.7 & Sparse & Full & Clear Skies\\
2005 April 05 & UH 2.2-m & 0.6 & Sparse & Full & Cirrus \\
2005 April 06 & UH 2.2-m & 0.6-0.8 & Sparse & Half & Cloudy\\
2005 April 07 & UH 2.2-m & 0.6 & Sparse & Half & Photometric\\
2005 April 09 & UH 2.2-m & 0.6-0.7 & Sparse & Half & Clear\\
2005 April 11 & UH 2.2-m & 0.6 & Sparse & Half & Clear\\
2005 April 12 & UH 2.2-m & 0.7 & Sparse & Half & Clear\\
2005 April 14 & UH 2.2-m & 0.7 & Sparse & Half & Clear\\
2005 April 15 & UH 2.2-m & 0.8 & Sparse & Half & Cloudy\\
2005 April 17 & UH 2.2-m & 0.8 & Dense & Half & Cloudy \\
2005 April 18 & UH 2.2-m & 0.8-1.0 & Dense & Half & Moon Rising \\
2006 February 01 & UH 2.2-m & 1.0 & Dense & Full & Focus Problems \\
2006 February 02 & UH 2.2-m & 0.6 & Dense & Full & Clear \\
2006 February 04 & UH 2.2-m & 1.5 & Dense & Full & Strong Winds \\
2006 February 24 & UH 2.2-m & 1.0-1.2 & Dense & Full & Windy \\
2006 April 24 & UH 2.2-m & 0.7 & Dense & Half & Cloudy/Clear \\
2006 April 29 & UH 2.2-m & 0.8 & Dense & Half & Clear,Windy \\
2006 April 30 & UH 2.2-m & 0.9 & Dense & Half & Clear,Windy \\
2006 May 01 & UH 2.2-m & 0.9-1.0 & Dense & Half & Windy\\
\enddata
\tablenotetext{a}{Sparse Sampling Survey or Follow-up Densely Sampled Lightcurves}
\end{deluxetable*}\label{table: obs_journal}

\begin{deluxetable*}{lcccccccc}
\tablecolumns{9}
\tablewidth{0pc}
\tabletypesize{\scriptsize}
\tablecaption{Photometry of Jovian Trojan Asteroids \label{binary_prob}}
\tablehead{
\colhead{Trojans} & \colhead{Tel} & \colhead{ $\bar m_{R}$\tablenotemark{a}} 
& \colhead{$m_1 - \bar m_{R}$\tablenotemark{b}} & \colhead{$m_2 - \bar m_{R}$\tablenotemark{b}} 
& \colhead{$m_3 - \bar m_{R}$\tablenotemark{b}} & \colhead{$m_4 - \bar m_{R}$\tablenotemark{b}} 
&\colhead{$m_5 - \bar m_{R}$\tablenotemark{b}} & \colhead{$\Delta m_{R}$\tablenotemark{c} } }
\startdata
884    & UH     & 16.37  & 0.13   & -0.09  & -0.07  & 0.08   & -0.05  & 0.22    \\
1172   & UH     & 15.78  & 0.06   & -0.05  & -0.04  & 0.03   &        & 0.11    \\
1173   & LOT    & 16.85  & 0.02   & -0.20  & -0.08  & 0.17   & 0.10   & 0.37    \\
1208   & UH     & 16.60  & 0.06   & 0.06   & -0.01  & 0.00   & -0.06  & 0.12    \\
1583   & UH     & 16.87  & -0.02  & -0.07  & 0.00   & 0.04   & 0.04   & 0.11    \\
1647   & UH     & 18.88  & -0.20  & -0.14  & 0.09   & 0.24   &        & 0.44    \\
1867   & UH     & 15.82  & 0.04   & 0.04   & 0.02   & -0.07  & -0.04  & 0.12    \\
1868   & UH     & 17.52  & 0.03   & -0.03  & -0.08  & 0.06   & 0.02   & 0.14    \\
1869   & UH     & 19.51  & -0.18  & 0.01   & 0.03   & 0.07   & 0.08   & 0.26    \\
1870   & UH     & 17.90  & -0.05  & -0.01  & 0.05   & -0.03  & 0.03   & 0.10    \\
1871   & UH     & 19.29  & 0.05   & 0.05   & 0.01   & -0.07  & -0.04  & 0.12    \\
1872   & LOT    & 17.99  & 0.09   & -0.03  & -0.01  & 0.01   & -0.06  & 0.15    \\
1873   & UH     & 17.24  & -0.14  & -0.05  & 0.11   & 0.08   &        & 0.25    \\
2146   & UH     & 17.79  & -0.07  & 0.07   & 0.05   & -0.06  & 0.00   & 0.14    \\
2207   & UH     & 16.03  & 0.05   & -0.02  & -0.03  & 0.03   & -0.03  & 0.08    \\
2241   & UH     & 15.95  & 0.11   & -0.15  & 0.01   & 0.03   &        & 0.26    \\
2260   & UH     & 17.47  & 0.03   & 0.12   & -0.09  & -0.03  & -0.03  & 0.22    \\
2357   & UH     & 15.93  & 0.01   & 0.04   & -0.02  & -0.03  &        & 0.07    \\
2357   & LOT    & 15.96  & -0.02  & -0.01  & -0.01  & 0.02   & 0.03   & 0.05    \\
2363   & UH     & 17.12  & 0.03   & 0.03   & -0.06  & 0.01   &        & 0.09    \\
2674   & LOT    & 16.54  & 0.19   & -0.23  & 0.03   & 0.26   & -0.23  & 0.49    \\
2893   & UH     & 16.62  & 0.14   & -0.03  & -0.11  & 0.00   &        & 0.26    \\
2895   & UH     & 17.24  & -0.01  & -0.04  & 0.08   & -0.02  &        & 0.12    \\
2895   & LOT    & 16.73  & -0.02  & 0.05   & 0.01   & -0.01  & -0.04  & 0.09    \\
2920   & UH     & 16.57  & 0.10   & 0.06   & -0.10  & -0.06  &        & 0.20    \\
3240   & UH     & 18.06  & -0.09  & -0.17  & 0.01   & -0.15  & 0.40   & 0.57    \\
3317   & UH     & 16.33  & 0.02   & 0.01   & -0.05  & -0.01  & 0.04   & 0.09    \\
3451   & UH     & 15.91  & -0.10  & 0.14   & 0.04   & -0.02  & -0.06  & 0.25    \\
3708   & UH     & 17.20  & 0.01   & -0.04  & -0.01  & 0.01   & 0.02   & 0.06    \\
3709   & UH     & 17.42  & -0.05  & -0.04  & 0.01   & -0.05  & 0.13   & 0.18    \\
4068   & UH     & 17.41  & -0.07  & -0.06  & 0.04   & 0.04   & 0.04   & 0.11    \\
4348   & UH     & 17.09  & 0.13   & 0.10   & -0.03  & -0.01  & -0.01  & 0.16    \\
4489   & LOT    & 17.04  & 0.08   & -0.01  & -0.06  & -0.01  &        & 0.13    \\
4707   & LOT    & 17.81  & -0.18  & 0.16   & -0.10  & -0.08  & 0.21   & 0.40    \\
4708   & LOT    & 17.35  & -0.20  & 0.13   & -0.04  & 0.11   &        & 0.33    \\
4709   & UH     & 15.92  & -0.05  & 0.05   & 0.09   & -0.05  & -0.06  & 0.15    \\
4715   & LOT    & 17.13  & 0.17   & -0.23  & -0.13  & 0.23   & -0.03  & 0.46    \\
4722   & LOT    & 17.28  & -0.02  & 0.00   & 0.01   & -0.04  & 0.05   & 0.08    \\
4754   & LOT    & 16.95  & 0.02   & 0.00   & 0.01   & -0.01  & -0.01  & 0.03    \\
4792   & UH     & 17.85  & 0.01   & 0.01   & 0.01   & -0.03  &        & 0.05    \\
4792   & LOT    & 17.56  & 0.17   & 0.03   & -0.10  & -0.06  & -0.04  & 0.27    \\
4805   & UH     & 17.73  & 0.01   & 0.04   & 0.04   & -0.09  &        & 0.14    \\
4827   & UH     & 17.86  & 0.01   & 0.07   & 0.02   & -0.06  & -0.05  & 0.13    \\
4828   & UH     & 17.63  & 0.13   & 0.11   & -0.06  & -0.19  &        & 0.32    \\
4828   & LOT    & 17.47  & 0.06   & 0.00   & -0.11  & 0.06   &        & 0.18    \\
4832   & LOT    & 17.55  & 0.01   & 0.00   & 0.01   & 0.00   & -0.02  & 0.03    \\
4833   & UH     & 17.25  & -0.18  & 0.10   & 0.13   & 0.05   & -0.10  & 0.31    \\
4834   & UH     & 17.70  & 0.06   & 0.02   & -0.02  & -0.04  & -0.03  & 0.10    \\
4867   & LOT    & 16.97  & 0.02   & -0.01  & -0.02  & -0.03  & 0.04   & 0.07    \\
5119   & UH     & 17.97  & 0.07   & 0.07   & -0.02  & -0.11  &        & 0.18    \\
5233   & UH     & 18.85  & 0.00   & -0.08  & 0.06   & 0.02   &        & 0.15    \\
5648   & UH     & 17.84  & 0.06   & 0.02   & -0.03  & -0.05  &        & 0.11    \\
6002   & UH     & 18.00  & 0.06   & 0.03   & -0.02  & -0.07  &        & 0.13    \\
9030   & UH     & 18.20  & -0.21  & 0.06   & 0.36   & -0.08  & -0.13  & 0.57    \\
9142   & LOT    & 18.19  & -0.08  & 0.05   & 0.04   & -0.01  & -0.01  & 0.13    \\
9431   & LOT    & 18.19  & 0.07   & -0.01  & -0.12  & -0.06  & 0.13   & 0.25    \\
9694   & UH     & 17.90  & -0.05  & -0.16  & -0.02  & 0.08   & 0.15   & 0.32    \\
11554  & LOT    & 17.31  & 0.03   & 0.00   & -0.03  & 0.00   & -0.01  & 0.06    \\
11668  & UH     & 19.33  & -0.05  & -0.02  & 0.14   & -0.03  & -0.08  & 0.22    \\
12649  & UH     & 19.64  & 0.04   & 0.00   & -0.06  & 0.00   & 0.02   & 0.10    \\
13402  & UH     & 19.08  & -0.02  & 0.00   & 0.00   & 0.02   & 0.01   & 0.04    \\
15527  & LOT    & 18.50  & 0.05   & 0.29   & -0.13  & -0.20  &        & 0.49    \\
16667  & UH     & 19.02  & -0.11  & 0.06   & 0.05   & 0.01   & 0.00   & 0.17    \\
17172  & LOT    & 17.83  & 0.04   & 0.03   & -0.04  & 0.00   & -0.03  & 0.07    \\
17365  & LOT    & 17.61  & -0.21  & 0.35   & 0.05   & -0.20  &        & 0.56    \\
17419  & UH     & 18.76  & -0.03  & 0.00   & 0.00   & 0.02   & 0.02   & 0.05    \\
17442  & UH     & 19.39  & 0.11   & 0.00   & 0.06   & -0.04  & -0.13  & 0.24    \\
17492  & UH     & 17.70  & 0.09   & 0.10   & 0.03   & -0.05  & -0.16  & 0.26    \\
18037  & UH     & 19.22  & -0.05  & -0.06  & -0.03  & -0.01  & 0.15   & 0.21    \\
18054  & UH     & 18.22  & -0.06  & 0.02   & -0.01  & -0.01  & 0.05   & 0.11    \\
23463  & UH     & 19.15  & -0.07  & 0.01   & 0.08   & -0.04  & 0.02   & 0.15    \\
23549  & UH     & 18.90  & -0.03  & 0.02   & 0.09   & 0.00   & -0.08  & 0.16    \\
24018  & UH     & 19.19  & 0.09   & 0.02   & -0.18  & -0.11  & 0.17   & 0.35    \\
24022  & UH     & 19.79  & 0.06   & -0.08  & -0.06  & 0.08   &        & 0.16    \\
24449  & UH     & 19.50  & 0.13   & 0.08   & -0.17  & -0.17  & 0.13   & 0.30    \\
24451  & UH     & 18.19  & 0.04   & 0.00   & 0.05   & -0.01  & -0.07  & 0.12    \\
24452  & UH     & 19.06  & -0.03  & 0.03   & -0.03  & 0.01   & 0.01   & 0.06    \\
24456  & UH     & 19.37  & -0.15  & 0.10   & 0.13   & 0.04   & -0.11  & 0.27    \\
24531  & LOT    & 19.72  & 0.25   & -0.07  & 0.05   & 0.00   & -0.23  & 0.48    \\
25344  & UH     & 19.22  & 0.13   & 0.01   & -0.13  & -0.11  & 0.09   & 0.26    \\
25347  & UH     & 19.23  & 0.09   & 0.20   & 0.04   & -0.16  & -0.17  & 0.37    \\
29314  & UH     & 19.44  & 0.22   & 0.31   & 0.21   & -0.21  & -0.53  & 0.83    \\
30498  & UH     & 19.59  & 0.00   & -0.07  & -0.12  & 0.10   & 0.09   & 0.22    \\
30499  & UH     & 19.76  & 0.05   & -0.03  & 0.04   & -0.07  & 0.01   & 0.12    \\
30505  & UH     & 19.02  & -0.13  & 0.15   & 0.08   & -0.22  & 0.12   & 0.34    \\
30506  & UH     & 18.78  & -0.19  & -0.18  & -0.02  & 0.19   & 0.20   & 0.39    \\
30704  & UH     & 18.67  & -0.08  & -0.03  & -0.01  & 0.11   &        & 0.19    \\
30942  & UH     & 18.52  & 0.04   & 0.02   & 0.00   & -0.02  & -0.04  & 0.08    \\
31806  & UH     & 19.51  & 0.15   & 0.07   & -0.09  & -0.03  & -0.10  & 0.25    \\
31814  & UH     & 19.81  & -0.11  & 0.11   & 0.23   & -0.09  & -0.16  & 0.39    \\
31819  & UH     & 18.90  & 0.20   & 0.01   & 0.00   & -0.03  & -0.17  & 0.37    \\
31820  & UH     & 20.06  & 0.16   & 0.09   & 0.05   & 0.12   & -0.40  & 0.56    \\
32482  & LOT    & 18.68  & 0.13   & -0.14  & 0.13   & 0.03   & -0.15  & 0.27    \\
32496  & UH     & 18.01  & 0.01   & -0.01  & -0.01  & 0.02   & -0.02  & 0.04    \\
32811  & UH     & 18.43  & -0.11  & -0.02  & 0.01   & 0.05   & 0.07   & 0.18    \\
47962  & UH     & 19.59  & 0.04   & -0.05  & -0.02  & 0.00   & 0.03   & 0.09    \\
51364  & UH     & 18.49  & 0.02   & 0.05   & 0.04   & -0.01  & -0.09  & 0.15    \\
53436  & UH     & 18.40  & -0.03  & 0.02   & 0.00   & 0.00   & 0.01   & 0.04    \\
55060  & LOT    & 18.85  & 0.27   & -0.09  & -0.22  & 0.03   &        & 0.48    \\
55419  & LOT    & 18.68  & 0.01   & -0.22  & -0.05  & 0.20   & 0.06   & 0.42    \\
65216  & UH     & 19.67  & 0.14   & -0.02  & -0.05  & -0.03  & -0.03  & 0.19    \\
67065  & UH     & 18.99  & 0.08   & -0.12  & -0.09  & 0.09   & 0.04   & 0.21    \\
69437  & UH     & 19.54  & -0.06  & 0.01   & 0.01   & 0.02   & 0.01   & 0.08    \\
73677  & UH     & 19.34  & 0.06   & 0.03   & 0.00   & -0.02  & -0.01  & 0.08    \\
85798  & UH     & 19.10  & -0.08  & 0.03   & 0.02   & 0.03   & 0.00   & 0.12    \\
1999 XJ55      & UH     & 19.29  & 0.04   & 0.00   & -0.03  & -0.01  &        & 0.06    \\
2000 TG61      & UH     & 19.76  & 0.01   & -0.01  & 0.00   & 0.02   & -0.03  & 0.04    \\
2000 SJ350     & UH     & 20.17  & -0.20  & -0.14  & -0.13  & 0.15   & 0.08   & 0.35    \\
2001 QZ113     & UH     & 19.53  & -0.02  & -0.02  & -0.02  & 0.00   & 0.05   & 0.07    \\
2001 XW71      & UH     & 20.24  & 0.06   & -0.03  & -0.05  & 0.17   & -0.08  & 0.24    \\
2001 QQ199     & UH     & 20.51  & -0.12  & -0.09  & 0.04   & 0.05   & 0.11   & 0.23    \\
2004 BV84      & UH     & 20.34  & 0.05   & -0.01  & 0.01   & -0.05  &        & 0.10    \\
2004 FX147     & UH     & 19.67  & 0.06   & -0.16  & -0.13  & 0.02   & 0.20   & 0.36    \\
2005 EJ133     & UH     & 20.15  & -0.11  & 0.01   & 0.00   & 0.08   & 0.01   & 0.18    \\
\enddata
\tablenotetext{a}{Mean R-Band Magnitude}
\tablenotetext{b}{R-Band Magnitude minus Mean R-Band Magnitude}
\tablenotetext{c}{Photometric Range}
\end{deluxetable*}

\begin{deluxetable*}{lcccccc}
\tablecolumns{7}
\tablewidth{0pc}
\tabletypesize{\scriptsize}
\tablecaption{Photometry of Jovian Trojan Asteroids \label{binary_prob}}
\tablehead{
\colhead{Trojan} & \colhead{$m_{R}(1,1,0)$\tablenotemark{a}} 
& \colhead{r [AU]\tablenotemark{b}} & \colhead{$\Delta$ [AU]\tablenotemark{c}} 
& \colhead{$\alpha$ [degrees]\tablenotemark{d}} & \colhead{D$_{e}$ [km]\tablenotemark{e}}
& \colhead{L4/L5}}
\startdata
884    & 8.53   & 5.66   & 5.34   & 9.9    & 146 & L5    \\
1172   & 8.00   & 5.68   & 5.24   & 9.4    & 193 & L5    \\
1173   & 9.08   & 6.02   & 5.27   & 6.6    & 150 & L5    \\
1208   & 8.86   & 5.69   & 5.17   & 9.0    & 134 & L5    \\
1583   & 9.30   & 5.33   & 4.92   & 10.2   & 99  & L4   \\
1647   & 11.50  & 5.20   & 4.70   & 10.1   & 37  & L4    \\
1867   & 8.18   & 5.34   & 5.12   & 10.7   & 163 & L5   \\
1868   & 9.91   & 5.50   & 5.00   & 9.5    & 80  & L4    \\
1869   & 12.10  & 5.49   & 4.75   & 7.5    & 34  & L4    \\
1870   & 10.29  & 5.42   & 5.03   & 10.1   & 64  & L5    \\
1871   & 11.47  & 5.46   & 5.46   & 10.5   & 36  & L5   \\
1872   & 10.78  & 5.51   & 4.61   & 4.7    & 84  & L5   \\
1873   & 9.71   & 5.11   & 5.02   & 11.3   & 78  & L5    \\
2146   & 9.98   & 5.69   & 5.28   & 9.6    & 77  & L4   \\
2207   & 8.73   & 5.05   & 4.61   & 10.7   & 127 & L5    \\
2241   & 8.34   & 5.17   & 5.17   & 11.1   & 148 & L5   \\
2260   & 9.92   & 5.39   & 4.92   & 9.8    & 77  & L4   \\
2357   & 8.34   & 5.29   & 4.90   & 10.4   & 149 & L5   \\
2357   & 8.79   & 5.29   & 4.51   & 7.1    & 164 & L5   \\
2363   & 9.74   & 5.24   & 4.69   & 9.7    & 85  & L5  \\
2674   & 9.37   & 5.17   & 4.49   & 8.6    & 111 & L5   \\
2893   & 8.75   & 5.56   & 5.50   & 10.4   & 128 & L5   \\
2895   & 9.79   & 5.25   & 4.69   & 9.6    & 81  & L5   \\
2895   & 9.67   & 5.24   & 4.39   & 6.3    & 118 & L5   \\
2920   & 9.23   & 5.25   & 4.64   & 9.3    & 111 & L4   \\
3240   & 10.04  & 5.92   & 5.61   & 9.5    & 75  & L5   \\
3317   & 8.44   & 5.78   & 5.39   & 9.5    & 157 & L5   \\
3451   & 8.38   & 5.44   & 4.90   & 9.4    & 163 & L5   \\
3708   & 9.29   & 5.93   & 5.41   & 8.6    & 113 & L5   \\
3709   & 9.77   & 5.58   & 5.04   & 9.1    & 87  & L4   \\
4068   & 9.97   & 5.33   & 4.78   & 9.5    & 78  & L4   \\
4348   & 9.51   & 5.49   & 4.95   & 9.2    & 97  & L5   \\
4489   & 9.26   & 5.54   & 5.37   & 10.3   & 104 & L4   \\
4707   & 10.60  & 5.53   & 4.62   & 4.4    & 96  & L5   \\
4708   & 10.05  & 5.34   & 4.65   & 8.2    & 84  & L5   \\
4709   & 8.53   & 5.30   & 4.71   & 9.3    & 153 & L5   \\
4715   & 9.85   & 5.30   & 4.62   & 8.4    & 91  & L5   \\
4722   & 10.04  & 5.44   & 4.60   & 6.0    & 102 & L5   \\
4754   & 10.04  & 5.22   & 4.29   & 4.1    & 129 & L5   \\
4792   & 10.00  & 5.69   & 5.25   & 9.5    & 74  & L5   \\
4792   & 10.11  & 5.68   & 4.86   & 6.1    & 98  & L5   \\
4805   & 10.06  & 5.46   & 5.11   & 10.2   & 71  & L5   \\
4827   & 10.51  & 5.08   & 4.70   & 10.9   & 55  & L5   \\
4828   & 10.18  & 4.96   & 4.81   & 11.6   & 59  & L5   \\
4828   & 10.37  & 4.96   & 4.40   & 10.1   & 63  & L5   \\
4832   & 10.00  & 5.94   & 5.03   & 4.2    & 128 & L5  \\
4833   & 9.58   & 5.61   & 5.07   & 9.1    & 95  & L4   \\
4834   & 9.80   & 5.94   & 5.38   & 8.4    & 91  & L4   \\
4867   & 9.86   & 5.20   & 4.43   & 7.4    & 97  & L5   \\
5119   & 10.08  & 5.74   & 5.30   & 9.3    & 72  & L5   \\
5233   & 11.32  & 5.05   & 4.92   & 11.4   & 35  & L5   \\
5648   & 9.76   & 5.88   & 5.62   & 9.7    & 82  & L5   \\
6002   & 10.34  & 5.55   & 4.97   & 8.9    & 66  & L5   \\
9030   & 11.03  & 5.11   & 4.46   & 9.1    & 49  & L5   \\
9142   & 10.41  & 5.84   & 5.27   & 8.4    & 70  & L5  \\
9431   & 10.51  & 5.52   & 5.17   & 10.0   & 59  & L4   \\
9694   & 10.75  & 5.39   & 4.51   & 5.5    & 78  & L4   \\
11554  & 10.12  & 5.32   & 4.53   & 6.9    & 90  & L5   \\
11668  & 11.74  & 5.87   & 5.04   & 5.9    & 47  & L4   \\
12649  & 11.61  & 5.90   & 5.58   & 9.5    & 36  & L5   \\
13402  & 11.20  & 5.72   & 5.35   & 9.6    & 43  & L5   \\
15527  & 10.95  & 5.32   & 5.01   & 10.5   & 47  & L4   \\
16667  & 10.85  & 6.17   & 5.88   & 9.1    & 54  & L5   \\
17172  & 10.59  & 5.45   & 4.61   & 6.0    & 80  & L5   \\
17365  & 10.31  & 5.54   & 4.69   & 5.8    & 92  & L5  \\
17419  & 11.33  & 5.38   & 4.81   & 9.3    & 43  & L5   \\
17442  & 11.62  & 5.43   & 5.35   & 10.6   & 34  & L5   \\
17492  & 10.10  & 5.42   & 5.07   & 10.3   & 70  & L5   \\
18037  & 11.50  & 5.51   & 5.21   & 10.3   & 37  & L5  \\
18054  & 10.85  & 5.19   & 4.74   & 10.3   & 50  & L5   \\
23463  & 11.57  & 5.27   & 5.05   & 10.9   & 34  & L5   \\
23549  & 11.54  & 5.10   & 4.76   & 11.0   & 35  & L5   \\
24018  & 11.65  & 5.44   & 4.95   & 9.7    & 36  & L5  \\
24022  & 12.12  & 5.66   & 5.12   & 9.0    & 30  & L5   \\
24449  & 11.96  & 5.36   & 4.94   & 10.2   & 30  & L5   \\
24451  & 10.33  & 5.89   & 5.39   & 8.8    & 70  & L5   \\
24452  & 11.78  & 5.01   & 4.63   & 11.1   & 31  & L5   \\
24456  & 11.86  & 5.33   & 4.90   & 10.2   & 31  & L5   \\
24531  & 11.79  & 5.76   & 5.57   & 9.9    & 33  & L4  \\
25344  & 11.54  & 5.62   & 5.11   & 9.2    & 39  & L5   \\
25347  & 11.44  & 5.57   & 5.32   & 10.2   & 38  & L5   \\
29314  & 11.84  & 5.46   & 5.02   & 9.9    & 32  & L5    \\
30498  & 11.78  & 5.70   & 5.33   & 9.7    & 34  & L5    \\
30499  & 12.16  & 5.32   & 5.06   & 10.7   & 26  & L5    \\
30505  & 11.60  & 5.32   & 4.76   & 9.5    & 37  & L5    \\
30506  & 11.06  & 5.43   & 5.24   & 10.6   & 44  & L5    \\
30704  & 11.20  & 5.34   & 4.85   & 9.8    & 43  & L5    \\
30942  & 11.20  & 5.17   & 4.64   & 10.0   & 43  & L5    \\
31806  & 11.73  & 5.67   & 5.29   & 9.7    & 34  & L5    \\
31814  & 12.16  & 5.65   & 5.10   & 8.9    & 30  & L5    \\
31819  & 11.65  & 5.14   & 4.57   & 9.8    & 36  & L5    \\
31820  & 12.46  & 5.50   & 5.03   & 9.6    & 25  & L5    \\
32482  & 11.36  & 5.26   & 4.66   & 9.2    & 42  & L5    \\
32496  & 10.30  & 5.63   & 5.17   & 9.5    & 68  & L5    \\
32811  & 11.14  & 5.00   & 4.64   & 11.2   & 41  & L5    \\
47962  & 12.04  & 5.54   & 4.95   & 8.9    & 32  & L5    \\
51364  & 11.42  & 4.95   & 4.34   & 9.9    & 39  & L5    \\
53436  & 11.36  & 5.21   & 4.35   & 6.4    & 54  & L4    \\
55060  & 11.41  & 5.34   & 4.83   & 9.6    & 40  & L5    \\
55419  & 11.12  & 5.51   & 4.98   & 9.1    & 47  & L5    \\
65216  & 12.49  & 5.44   & 4.54   & 5.2    & 36  & L4    \\
67065  & 12.02  & 5.20   & 4.30   & 5.3    & 44  & L4    \\
69437  & 11.89  & 5.55   & 5.10   & 9.7    & 32  & L5    \\
73677  & 11.99  & 5.27   & 4.70   & 9.6    & 31  & L5    \\
85798  & 11.89  & 5.45   & 4.57   & 5.6    & 45  & L4    \\
1999 XJ55      & 12.21  & 5.29   & 4.42   & 6.0    & 38 & L4    \\
2000 TG61      & 12.23  & 5.47   & 4.92   & 9.2    & 28 & L5   \\
2000 SJ350     & 12.55  & 5.44   & 5.03   & 10.0   & 23 & L5     \\
2001 QZ113     & 11.98  & 5.39   & 4.98   & 10.2   & 30 & L5     \\
2001 XW71      & 12.71  & 5.51   & 4.89   & 8.7    & 23 & L5     \\
2001 QQ199     & 12.59  & 6.36   & 5.48   & 4.6    & 37 & L5     \\
2004 BV84      & 12.95  & 5.37   & 4.74   & 8.8    & 21 & L5     \\
2004 FX147     & 12.61  & 5.25   & 4.39   & 6.0    & 31 & L4     \\
2005 EJ133     & 12.72  & 5.39   & 4.80   & 9.1    & 23 & L5     \\
\enddata
\tablenotetext{a}{Absolute Magnitude (see Equation 1)}
\tablenotetext{b}{Heliocentric Distance}
\tablenotetext{c}{Geocentric Distance}
\tablenotetext{d}{Phase Angle}
\tablenotetext{e}{Effective Diameter (see Equation 2)}
\end{deluxetable*}

Raw data frames were bias subtracted, then flat fielded using a
master flat field produced from median filtering dithered images
of the sky taken at dusk and dawn.  Landolt (1992) standard star
fields were imaged and measured to convert the instrumental
magnitudes to an absolute magnitude scale.  An aperture radius of
eight pixels was consistently used throughout the observations
for images taken on both telescopes.  Median sky values were 
determined using an adjacent annulus around the aperture having 
an outer radius of 20 pixels.  The reason for similar aperture
and sky annulus size on both telescopes, despite
differing pixel scales was because of the significantly worse
seeing conditions at Lulin (see Table 1).  
For the sparse sampling survey, two images were taken in each 
setting and then averaged to obtain the brightness measurement.
The photometric uncertainties are small ($\le$ 0.02 mag.) compared to
the photometric variability that is the subject of interest and so we
have ignored these uncertainties in our
presentation of the data.
For the densely sampled lightcurves, errors for each observation
were calculated using Poisson statistics.
The instrumental magnitude of the 
asteroid in each image was subtracted from the brightness of a 
nearby field star.  The field star was chosen to be persistent
in all five observations and helped reduce photometric errors by 
providing a correction for weather variations occuring throughout
the night.  Images in which the asteroid was affected by proximity to a
field star were rejected and resulted in some Trojans 
having only four measurements of brightness rather than five.

\section{Results}
Tables 2 and 3 contain results of the sparsely sampled lightcurve survey.
In Table 2, the average R band magnitude,
$\overline{m_R}$ is listed, along with the independent measurements of the
asteroid's brightness, expressed as deviations from the mean 
magnitude.  The last column shows the maximum deviation measured,
which gives a lower limit to the photometric range of each asteroid.
Table 3 contains the absolute magnitude, $m_R(1,1,0)$, which is
defined as the magnitude an object would have if placed at 
heliocentric ($r$) and geocentric $(\Delta)$ distances 
of 1 AU, and at a phase angle of $ \alpha $ = 0 degrees. The
conversion between the apparent magnitude, $m_R$ and absolute
magnitude, $m_R(1,1,0)$ is

\begin{equation} \label{equ: abs_mag}
m_R(1,1,0) = m_R - 5log(r\Delta) - \beta \alpha ,
\end{equation}

where $ \beta $ is the phase coefficient for which we used a value of 
0.04 magnitudes per degree for the low albedo Trojan asteroids \citep{bowell}.  
Also listed in Table 3 is an estimate of the equivalent 
circular diameter, $D_e$ which was calculated using \citep{russell}

\begin{equation}
m_R(1,1,0)=m_{\sun}-2.5\log\left[ \frac{p D_e^2}{4 \times 2.25 \times 10^{16} } \right].
\end{equation}

\begin{figure}
\epsscale{1.0}
\plotone{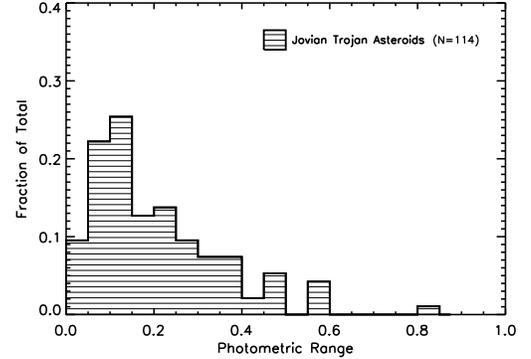}
\caption{Histogram of the distribution of photometric ranges found from 
sparse-sampled observations of 114 Jovian Trojan asteroids.}
\end{figure}

\begin{figure}
\epsscale{1.0}
\plotone{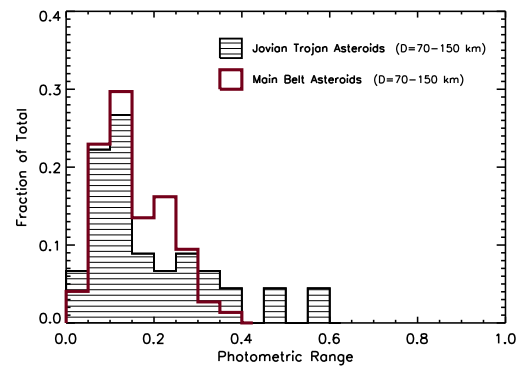}
\caption{Histogram of the photometric ranges of Jovian Trojan asteroids
and Main Belt asteroids with diameters between 70-km and 150-km.
Data for Main Belt asteroids taken from Barucci et al. (2002).}
\end{figure}

Here, \textit{p} is the geometric albedo, for which a value of 0.04 was
used throughout \citep{fernandez} and $m_{\sun}$ = -27.1 is the 
apparent red magnitude of the sun \citep{cox}.

Figures 4 shows the distribution of photometric
ranges shown by the Trojan asteroids in the sparsely-sampled
lightcurve survey.  For comparison, Figure 5 shows the 
photometric range distributions of both the Trojan and
Main Belt asteroids with diameters between 70-km 
and 150-km (Main Belt asteroid data taken from \citet{barucci}).
Figure 5 reveals that a larger fraction of Trojan
asteroids have photometric ranges larger than Main Belt
asteroids, similar to previous studies by \citet{hartmann88}.
A Kolmogorov-Smirnov statistical test found a 32.1\%
probability that the two distributions are
drawn from the same parent distribution.

\begin{figure}
\epsscale{1.0}
\plotone{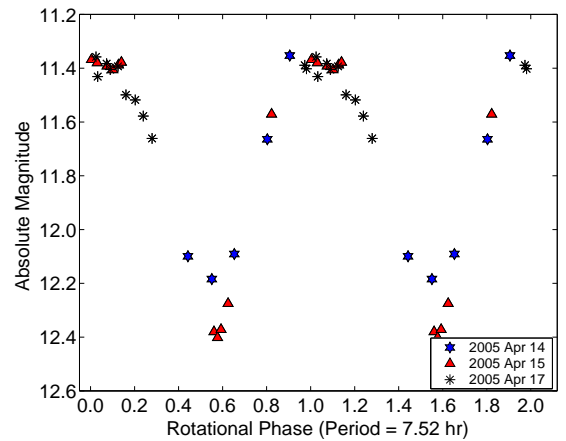}
\caption{Absolute magnitude (calculated from equation 1) of Trojan 
asteroid (29314) in April 2005.  Data are phased to a single-peaked
lightcurve period of 7.52 hours.}
\end{figure}

\begin{figure}
\epsscale{1.0}
\plotone{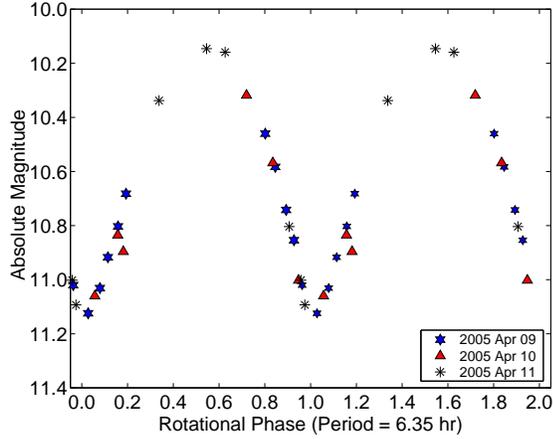}
\caption{Absolute magnitude (see equation 1) of Trojan 
asteroid (17365) in April 2005.  Data are phased to a
single-peaked lightcurve period of 6.35 hours.}
\end{figure}

\begin{figure}
\epsscale{1.0}
\plotone{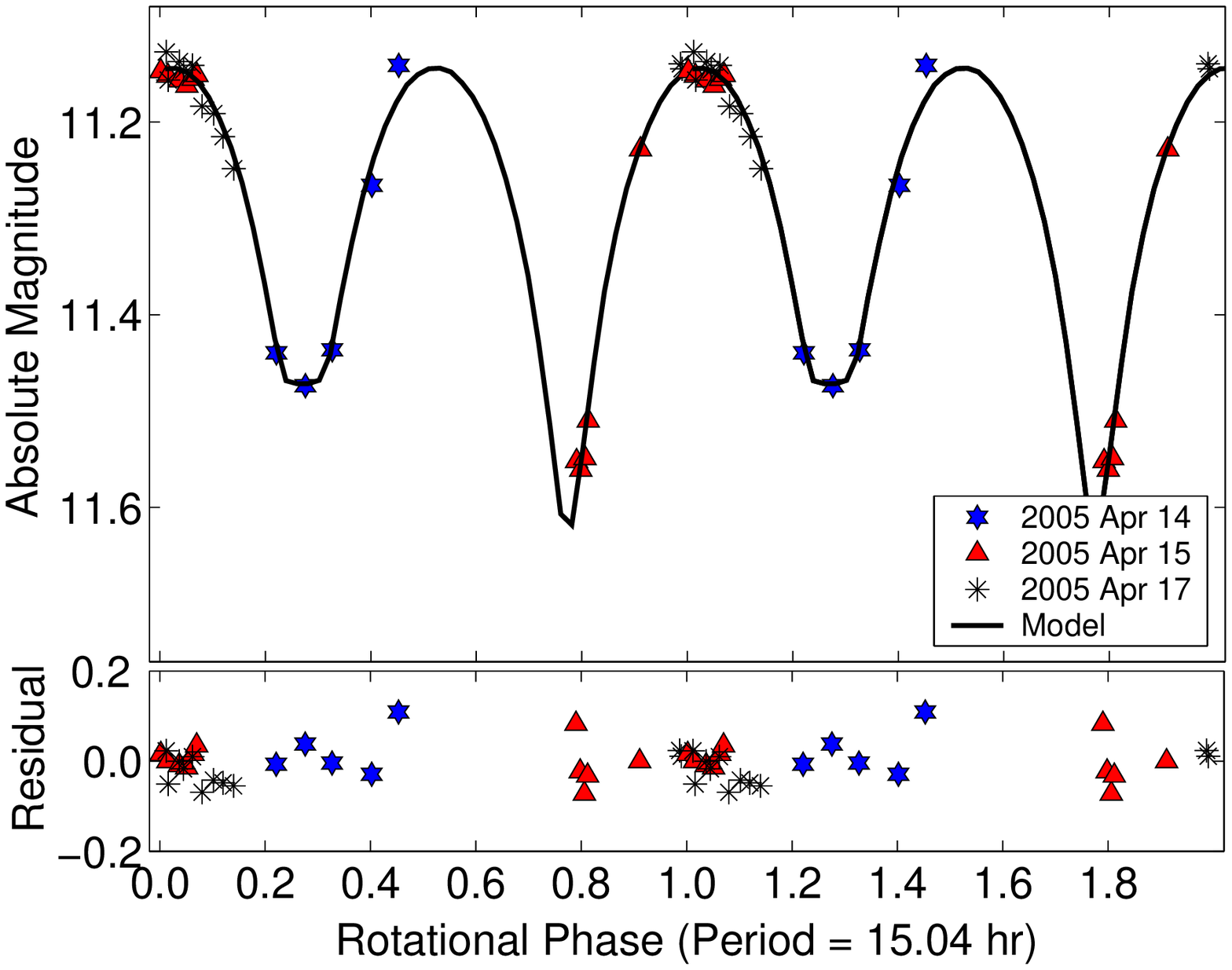}
\caption{Absolute magnitude (see equation 1) of Trojan asteroid
(29314) in April 2005.  Data phased to a double-peaked lightcurve
period of 15.04 hours.  Best fit Roche binary equilibrium model 
is overplotted. }
\end{figure}

\begin{figure}
\epsscale{1.0}
\plotone{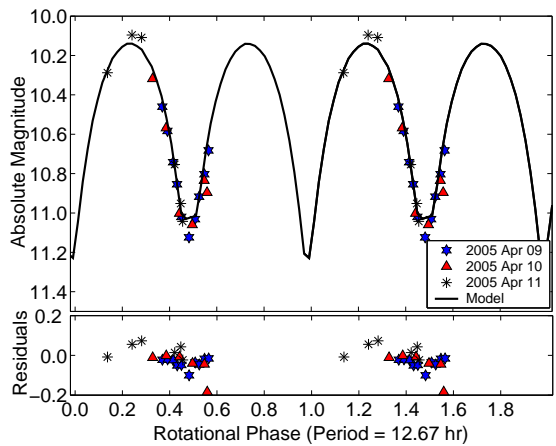}
\caption{Absolute magnitude (see equation 1) of Trojan asteroid
(17365) in April 2005.  Data phased to a double-peaked lightcurve
period of 12.67 hours.  Best fit Roche binary equilibrium model 
is overplotted. }
\end{figure}

Trojan asteroids (17365) and (29314) showed the largest photometric
ranges in the sparsely-sampled photometry, 
with 0.56 $\pm$ 0.02 magnitudes and 0.83 $\pm$ 0.03 magnitudes, 
respectively (see Table 2).  Follow-up observations to obtain densely
sampled optical lightcurves for both Trojan asteroids were taken
using the University of Hawaii 2.2-m telescope between 2005 April
9th and 17th.  We were unable to complete the observations
due to bad weather coupled with the fact the asteroids were quickly
setting.  We were however, able to confirm the large photometric
ranges to motivate further study of these Trojan asteroids (see Figures 6 through 9).  
In our first dense light curve study, in 2005, 
asteroid (17365) had a photometric range of 0.98 $\pm$ 0.02 
magnitudes, centered at a mean of 10.64 $\pm$ 0.01 magnitudes, 
while asteroid (29314) had a peak-to-peak lightcurve
amplitude of 1.05 $\pm$ 0.03 centered on 11.89 $\pm$ 0.02
magnitudes.

\begin{figure}
\epsscale{1.0}
\plotone{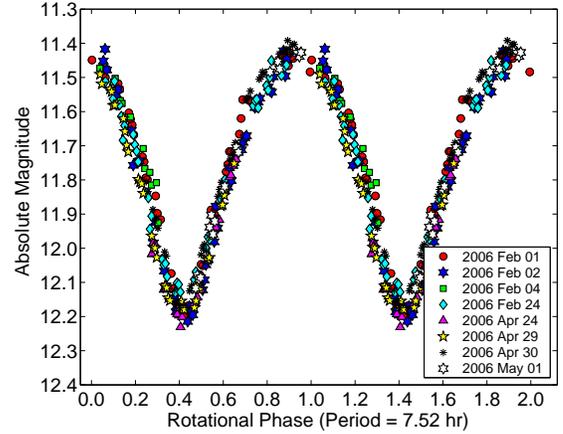}
\caption{Absolute magnitude (see equation 1) of Trojan asteroid 
(29314) between February and May 2006.  
Data are phased to a single-peaked lightcurve period of 7.52 hours.}
\end{figure}

\begin{figure}
\epsscale{1.0}
\plotone{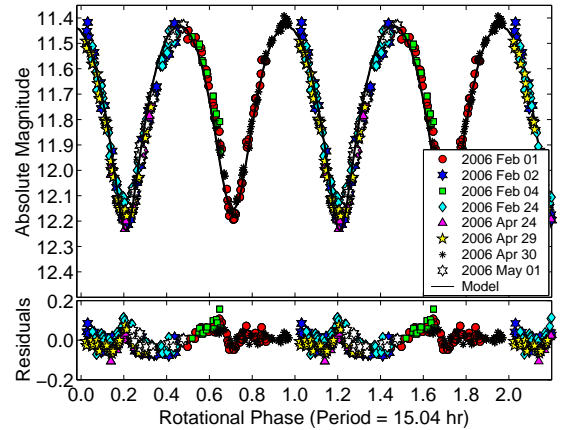}
\caption{Absolute magnitude (see equation 1) of Trojan asteroid 
(29314) between February and May 2006.  Data are phased to a double-peaked 
lightcurve period of 15.04 hours.  Best fit Roche binary equilibrium model 
is overplotted. } 
\end{figure}

\begin{figure}
\epsscale{1.0}
\plotone{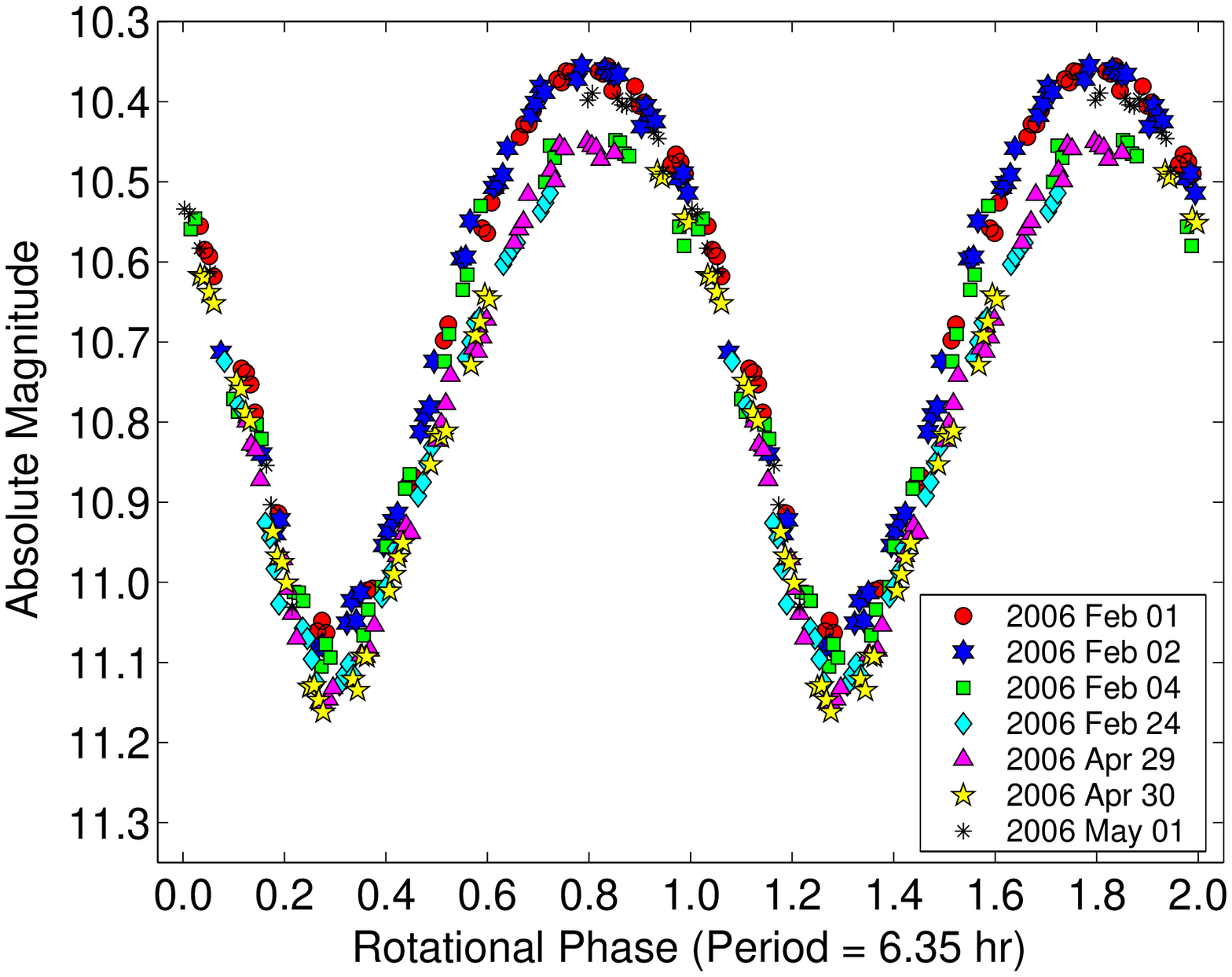}
\caption{Absolute magnitude (see equation 1) of Trojan asteroid (17365) 
between February and May 2006.  Data are phased to a single-peaked lightcurve 
period of 6.35 hours.  }
\end{figure}

\begin{figure}
\epsscale{1.0}
\plotone{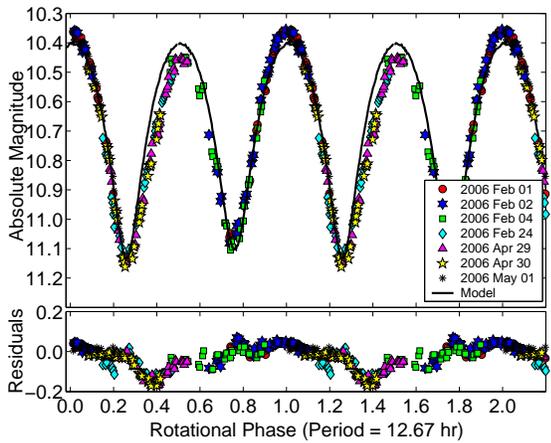}
\caption{Absolute magnitude (see equation 1) of Trojan asteroid (17365)
between February and May 2006.  Data are phased to a double-peaked 
lightcurve period of 12.67 hours.  Best fit Roche binary equilibrium 
model is overplotted. }
\end{figure}

To complete the lightcurve study, we 
continued optical observations of both candidate
contact binary asteroids in 2006.  Figures 10 through 13 show the
results of the photometric observations.  In 2006, asteroid
(17365) showed a photometric range of 0.81 $\pm$ 0.02 magnitudes,
centered at a mean absolute magnitude of 10.76 $\pm$ 0.01.  Asteroid 
(29314) shows a peak-to-peak amplitude of 0.86 $\pm$ 0.03 magnitudes, 
with a mean absolute magnitude of 11.80 $\pm$ 0.02.

The phase dispersion minimization (PDM) method \citep{stellingwerf}
was used to determine possible rotation periods for each asteroid.  
Figures 14 and 15 show plots of $\Theta$, which characterizes the dispersion
in the data phased to a given period (see 
\citet{stellingwerf} for more information).  The most likely rotation 
periods correspond to the smallest values of theta.  Several periods appeared to 
minimize theta, but when used to phase the data, 
the results were not persuasive lightcurves.  In fact, only two 
periods per asteroid produced convincing lightcurve results.
For Trojan (29314), minima consistent with the data occur at
periods of 0.3133 $\pm$ 0.0003 days (7.518 $\pm$ 0.007 hr), and a 
double-peaked period of 0.6265 $\pm$ 0.0003 days 
(15.035 $\pm$ 0.007 hr).  Asteroid (17365) shows a single-peaked
lightcurve period of 0.2640 $\pm$ 0.0004 days (6.336 $\pm$ 0.009 hr)
and double-peaked period of 0.52799 $\pm$ 0.0008 days 
(12.672 $\pm$ 0.019 hr).

While both the single-peaked and double-peaked periods produce good 
fits for Trojan asteroid (29314), the double-peaked lightcurve 
is more convincing. The lightcurve of (29314) shows subtle differences in
the shapes of the two minima, which is obvious by the spread in the data
when phased to the single-peaked period (see Figure 10 and 11).
Asteroid (17365) shows a more obvious double-peaked lightcurve 
(see Figures 12 and 13) with maxima of different shapes.  
The maxima of (17365) differ by 0.10 $\pm$ 0.01 magnitudes while 
the minima differ by 0.06 $\pm$ 0.01 magnitudes.

\begin{figure}
\epsscale{1.0}
\plotone{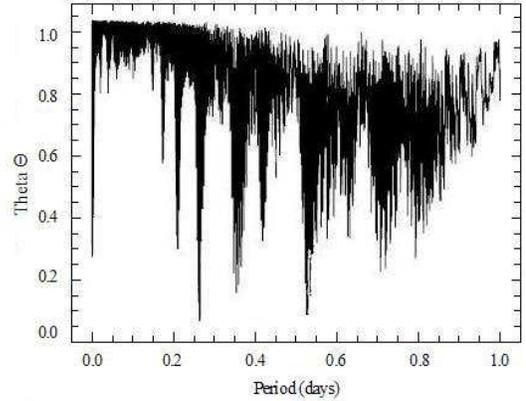}
\caption{Phase Dispersion Minimization (PDM) plot for Trojan
asteroid (17365) showing $\Theta$ versus period.  
Probable periods are at minimum $\Theta$ values: 0.2640 $\pm$ 0.0004 days
and 0.52799 $\pm$ 0.0008 days.}
\end{figure}

\begin{figure}
\epsscale{1.0}
\plotone{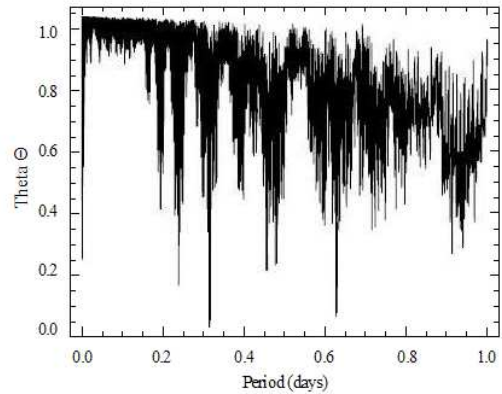}
\caption{Phase Dispersion Minimization (PDM) plot for Trojan
asteroid (29314) showing $\Theta$ versus period.
Probable periods minimize $\Theta$:  0.3133 $\pm$ 0.0003 days
and 0.6265 $\pm$ 0.0003 days.}
\end{figure}

\subsection{Candidate Contact Binary Asteroids}
Trojan asteroids (17365) and (29314) show strong evidence of being contact 
binaries.  Both asteroids reveal photometric ranges greater than 
0.9 magnitudes, sufficiently long rotation periods ($<$ 2 rotations per day)
and lightcurve profiles (qualitatively similar to 624 Hektor) 
containing U-shaped maxima and 
V-shaped minima.  Here, we speculate about all possible explanations 
for the brightness variations in the lightcurve observations of these 
Trojan asteroids, including albedo variations, elongated shapes or binarity
\citep{dunlap,cook71,hartmann78,weidenschilling80}.

Surface albedo contrasts provide a possible but
unconvincing explanation for the large brightness variations of the Trojans.  
Amongst Solar system objects, only Iapetus, a satellite of Saturn, shows strong
spatial albedo variations which account for its large lightcurve amplitude.  
However, Iapetus'  synchronous rotation about Saturn plays a large role in 
producing the dichotomous behaviour of the satellite \citep{cook70} and
this circumstance is not relevant in the context of the Trojan 
asteroids.  Amongst previously studied asteroids, double-peaked lightcurves
are almost always caused by rotational variations in the projected area, and
reflect the elongated shapes of the bodies.
While albedo contrasts cannot be formally ruled out, 
we feel that they are an unlikely cause of the observed brightness
variations.

Increasing evidence suggests asteroids have little or
no internal strength, probably as a result of impacts that disrupt but do
not disperse the object \citep{farinella81,pravec}. The Trojan
asteroids have undergone a collisional history that is either
similar to that of the main-belt asteroids \citep{marzari}
or perhaps even more intense \citep{davis,barucci},
making it highly probable that they, too, are gravity dominated
``rubble piles", strengthless or nearly so in tension \citep{farinella81}.
Studies have found that only the smallest main-belt asteroids, 
with diameters less than 0.15-km, have sufficient internal 
strength to overcome gravity \citep{pravec}.
Figure 5 from \citet{pravec} shows observations of decreasing
maximum spin rate with increasing lightcurve amplitude (a proxy for 
elongation) of near-earth asteroids.  This observation indicates a lack
of fast rotating elongated bodies, which implies that
asteroids larger than $\sim$ 0.15-km
are structurally weak and lack the tensile strength to 
withstand high rotation rates without becoming unstable and flying apart.
Also evident in Figure 5 \citep{pravec} is the tendency of
fast rotators to have spheroidal shapes, an indicator of
gravity-dominated bodies  which do not possess the internal 
strength to resist gravity.  Collectively, the observations point 
to asteroids being bodies of negligible strength, whose
shapes are dominated by rotation and gravity.

Rotation rates must lie between 4 and 6 rotations per day 
in order for rotational elongation of a structurally weak body 
to be maintained.  This is the range for which Jacobi ellipsoids 
are possible figures of equilibrium \citep{leone,farinella97}.
If the rotation rate was much higher than 6 rotations per day, 
the body would fall apart, while at a much lower rotation rate,
the body would adopt a spherical figure of equilibrium.
In 2005, both asteroids (17365) and (29314) showed photometric
variations larger than 0.9 magnitudes, above the threshold
for rotational instability in a structurally weak body.
Additionally, both asteroids have double-peaked lightcurve
periods that are too slow to cause sufficient rotational
elongation.  Both observations indicate that rotationally-induced
elongation is an insufficient explanation for 
the brightness variations of these Trojan asteroids.

\begin{figure}
\epsscale{1.0}
\plotone{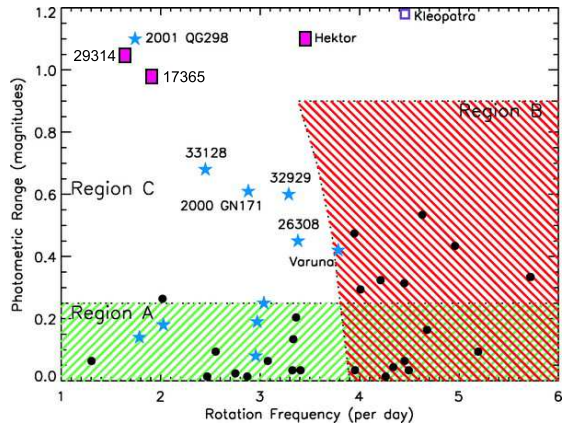}
\caption{Modification of Figure 5 from Sheppard \& Jewitt, 2004
(originally taken from Leone et al. 1984) to include
contact binary candidates (17365) and (29314).  Stars
represent Kuiper Belt objects, black circles are main-belt
asteroids with diameters larger than 50-km and pink squares are 
the candidate binary Trojans (17365), (29314) and 624 Hektor.  Region
A includes all objects whose photometric range could be caused
by albedo, elongation or binarity.  Region B contains objects that
are likely to be rotationally elongated.  Only binaries are expected
in Region C.}
\end{figure}

We are therefore left with the strong possibility that
Trojan asteroids (29314) and (17365) are contact binaries.  
Figure 16 is a plot of rotation periods and photometric ranges 
of several well studied Kuiper Belt objects and main-belt 
asteroids.  It is divided into three main regions:  
Region A spans the photometric ranges that can be explained by 
albedo variations, elongation or binarity of an asteroid.  
Region B represents the characteristics 
explained by albedo variations or rotational elongation of an 
object, while variations in region C can only be explained by binary asteroids.
Both Trojan asteroids lie well within Region C, alongside
contact binaries 216 Kleopatra, 624 Hektor and 2001 QG$_{298}$, 
contributing to their suspected binary nature.

\begin{figure}
\epsscale{1.0}
\plotone{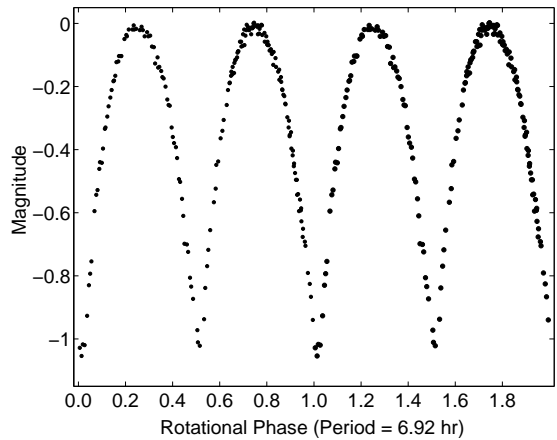}
\caption{Lightcurve of 624 Hektor in April 1968, taken from \citet{dunlap}.
Note the similarities between lightcurves of (29314), (17365) and 624 Hektor.}
\end{figure}

The lightcurve of a contact binary is expected to show
U-shaped or spread out maxima and V-shaped or notched minima, 
as shown by the lightcurves of 2001 QG$_{298}$ (see \citet{sheppard}) 
and 624 Hektor (see Figure 17).  These characteristic
lightcurve profiles are unlike the 
distinctive ``notched" profile expected for wide, eclipsing binaries 
which remain flat for the majority of the orbit, and contain sharp 
dips during the relatively short eclipsing events.  
The photometric observations of Trojan asteroids (29314) and (17365) are
consistent with lightcurve profiles expected of very close
or contact binary systems.

624 Hektor was recently discovered to possess a satellite of 
diameter 15-km using Keck Laser Guide Star Adaptive Optics 
\citep{marchis_iauc}, but an independent density estimate derived from the
orbital motion of this satellite has not yet been published.  Additionally, the imaging observations of 624 
Hektor indicate that its primary component has a double-lobed nature. 
Similarities are obvious between the lightcurves
of 624 Hektor, (17365) and (29314) (see Figures 11, 13 and 17)
and consistent with our interpretation that the latter two asteroids are
contact binaries.

We used equilibrium models of Roche binaries to determine how well the
photometric observations of (17365) and (29314) could be matched by theoretical lightcurves
of contact binary systems.  A Roche binary consists of a pair of homogeneous
bodies in hydrostatic equilibrium orbiting each other.  A strength of this
modeling is the ability to estimate densities for the asteroids without knowing
the sizes of the binary components.  The exact shapes and rotation rates of the
Roche binaries were calculated using the mathematical description presented in
\citet{leone} (see also \citet{chandrasekhar}). Binary configurations were
calculated for secondary to primary mass ratios from $q=0.25$ to $q=1.00$ in
steps of 0.01. For each value $q$, Equations 1 to 3 of \citet{leone} were
solved simultaneously to find possible shapes and orbital frequencies for the
primary. The same equations were then solved using mass ratio $q'=1/q$ to
calculate the shapes and orbital rates for the secondary.  Finally, valid
binaries are uniquely selected by matching pairs (q,1/q) with the same orbital
frequency. This procedure is described in detail in \citet{leone} and
\citet{lacerda}. 

The models were ray-traced using publicly available software POV-Ray
(http://www.povray.org), but the surface scattering routine of POV-Ray was
rewritten to allow better control of the scattering function. The scattering
law used here was first implemented by \citet{kaas01}.
It linearly combines single (Lommel-Seeliger) and multiple (Lambert) scattering
terms using a parameter $k$ \citep{takahashi}, which varies from 0 to 1.
The resulting reflectance function is 
\begin{equation}
r\propto(1-k)\,\frac{\mu_0}{\mu_0+\mu}+k\,\mu_0 
\end{equation} 
where $\mu_0$ and $\mu$ are the cosines of the incidence and emission angles.
When $k=0$, only single scattering is present, while $k=1$ simulates pure
multiple scattering of light off the surface of the binaries. All binary
configurations were raytraced for $k$ between 0 and 1 in steps of 0.1. Two
viewing geometries were modelled, at aspect angles of 75 and 90$\,$deg
(equator-on).  The aspect angle lies between the line of sight of the
observations and the rotation axis of the body.  Simulated illumination angles
were chosen to match the phase angles at the time the data were taken. In
total, nearly 50000 models were computed for comparison with the data.

Observations of (17365) and (29314) were simultaneously fitted
for the different viewing orientations in 2005 and 2006 to find
the best shape interpretation for the asteroids. 
We assumed that the objects were viewed equatorially in 2005, thus
producing the larger photometric range in the discovery epoch data.
This assumption was encouraged by the fact that an aspect angle of 
75 degrees (rather than 90 degrees) produced a better fit with the 
2006 observations (see Figures 10 and 12). 

Figures 8, 9, 11 and 13 show the best-fit models overlaying lightcurve data,
with residuals plotted underneath.  Best fit models were found 
by minimizing chi-squared.  Small deviations ($\sim$ 0.1 magnitudes) from the 
binary model are evident for both asteroids, but are negligible compared 
with the total range of the observations, the more important parameter.
Presumably, the deviations are caused by irregularities on the surface
of the asteroids, which were not included in the simple binary model,
but without which the asteroids would be considered odd.  
The ability of the models to simultaneously fit two 
epochs of photometric observations lends strong support to the idea
that we observed contact binary asteroids over two years at different
viewing geometries.

\begin{figure}
\epsscale{1.0}
\plotone{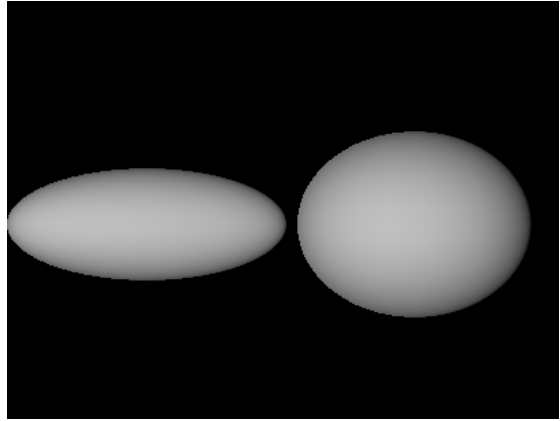}
\caption{Shape interpretation of Trojan asteroid (29314)
from Roche binary equilibrium models.}
\end{figure}

\begin{figure}
\epsscale{1.0}
\plotone{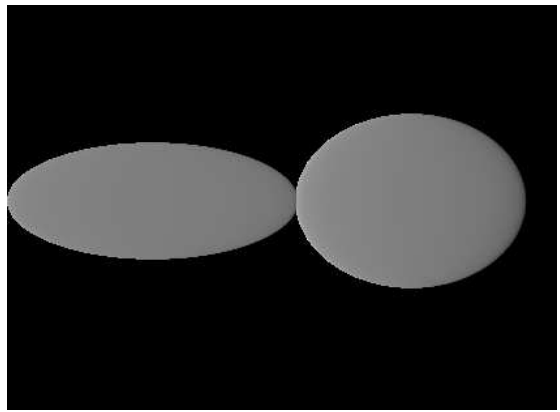}
\caption{Shape interpretation of Trojan asteroid  
(17365) from Roche binary equilibrium models.}
\end{figure}

Figures 18 and 19 show the shapes derived from the binary models for (17365) and
(29314).  Orbital periods combined with shape information allowed us to
estimate the densities of the asteroids.  The components of our model of
asteroid (29314) were found to have a mass ratio of $0.4^{+0.5}_{-0.1}$ and
a density of $590^{+40}_{-80}$ kg/m$^3$, while our best model of asteroid
(17365) has a mass ratio of $0.6^{+0.2}_{-0.1}$ and density of
$780^{+50}_{-80}$ kg/m$^3$.  These low densities suggest porous asteroid
interiors.  If (29314) and (17365) have a rock/ice composition similar to the
moons of Jupiter, (29314) would have a porosity of $\sim$ 64\%, while (17365)
would have a smaller porosity of 50\% (see Figure 3 from \citet{marchis}).  If
(17365) and (29314) were composed purely of water ice, their porosities would be
15\% and 35\%, respectively \citep{marchis}.  This pure water ice composition
is unrealistic, however.  It is interesting to note that our low density
measurements are consistent with 617 Patroclus \citep{marchis}.

Among the Trojans, only 624 Hektor is known to have a comparable
lightcurve amplitude, making (29314) and (17265) the 2nd and 3rd known
Trojans to show such large rotational variations.  
Lightcurve analysis suffers from the notorious non-uniqueness problem, which
arises from the ability to reproduce any lightcurve with a complicated pattern
of surface markings and shapes.  Our interpretation is not unique, but is the
simplest, most plausible explanation for the behaviour of the Trojan asteroids.

\section{Discussion: Binary Fraction}
Following the method outlined in \citet{sheppard} to account for
the geometrical circumstances of the observations, we were able to
estimate the fraction of contact binary systems among the Jovian Trojan
asteroids.  This method uses two very crude approximations.
In the first approximation, the binary system is simplified to 
be an elongated, rectangular object with dimensions 
a $\geq$ b = c, having a lightcurve amplitude as follows:

\begin{equation} 
\Delta m = 2.5\log\left\{\frac{1+\tan\theta}{\frac{b}{a} + \tan\theta}\right\}.
\end{equation}

The range of lightcurve amplitudes used to identify contact binary asteroids is
0.9 to 1.2 magnitudes.  For the maximum amplitude of 1.2 magnitudes and viewing
angle of $\theta$ = 0$\degr$, an axis ratio of $\frac{a}{b}$ = 3 is calculated
from Equation 4.  Using this axis ratio and the minimum expected amplitude of 0.9
magnitudes, a viewing angle of 10$\degr$ was determined.  Therefore, the range
of lightcurve amplitudes expected for a contact binary asteroid would only
be observed if the Earth lies within 10$\degr$ of the equator of the asteroid.  
The probability that the Earth would lie within 10$\degr$ of the equator
of a randomly oriented asteroid is P($\theta \leqslant 10\degr$) = 0.17.  
We found two suspected contact binary asteroids in our sample of 114
Trojan asteroids, so the fraction of contact binary Jovian Trojan
asteroids is approximately $\frac{2}{114(0.17)}$ = 10 \%.

A second approximation uses an ellipsoid shape to represent the contact 
binary asteroid, again having dimensions a $\geq$ b = c, and having a 
lightcurve amplitude expressed by the following:

\begin{equation} 
\Delta m = 2.5\log\left(\frac{a}{b}\right) - 1.25\log\left\{\left[\left(\frac{a}{b}\right)^2 - 1\right]\sin^2 \theta + 1 \right\}.
\end{equation}

Using the axis ratio of $\frac{a}{b}$ = 3, in order to observe 
photometric ranges between 0.9 and 1.2 magnitudes, the Earth must
lie within 17$\degr$ of the equator of the ellipsoidal asteroid.
The probability of a randomly oriented object having this 
geometrical orientation relative to the observer is 
P($\theta \leqslant 17\degr$) = 0.29, implying a contact binary 
fraction of $\frac{2}{114(0.29)}$ = 6 \%.

We conclude that the fraction of contact binary Trojan asteroids
is $\sim$6\% to $\sim$10\%.  This is a lower limit to the actual fraction as some of 
the objects not found in the survey sample to have large 
amplitudes might in fact have them because the sparse sampling
method is not 100\% efficient.  The existence of likely contact
binary 624 Hektor separately suggests that the binary fraction is high.

Binaries with equal-sized components are rare in the main-belt (the 
frequency of large main-belt binaries is $\sim$ 2\% \citep{richardson06}) and have
yet to be observed in the near-earth asteroid population.  However, they
are abundant in the observed binary Kuiper Belt population, where the 
fraction lies between 10\% and 20\% \citep{sheppard}.  The results of this study
show that there are three Jovian Trojan asteroids that reside in Region
C.  The observations tend to suggest a closer relationship between
the binary populations of the Kuiper Belt and the Trojan clouds.  
This correlation could signify similar binary formation mechanisms
in the two populations.  This is an interesting connection considering
that in one model of formation, the Trojans are actually captured Kuiper
Belt objects \citep{morbidelli}.
However, it is clear that the total binary fractions in the Kuiper Belt
and in the Trojans needs to be more tightly constrained
before conclusions can be made.

The contact binaries detected were skewed towards those with components of 
comparable sizes, which are capable of producing photometric ranges $\geq$ 0.9 
magnitudes.  For mass ratios $\ll$ 1, sparse sampling would more likely 
miss the eclipsing event and the photometric range would be $\leq$ 0.9 magnitudes 
and would not attract our attention.
The method was strongly dependent on geometrical circumstances, and only
binaries viewed edge-on or almost equatorially would be detected in our survey.
Additionally, sparse sampling is only able to put lower limits on the
photometric range of an asteroid, making the binary fraction a lower limit
estimate.  Only binaries with sufficiently short orbital periods 
(optimally between 6 to 12 hour rotation
periods) would be detected, so wide binaries were not accounted for in this study.
Therefore, again the measured binary fraction is a strong lower limit 
to the actual fraction and is suggestive of a
significant binary population among the Trojan clouds.  

Our project is a pilot study for the much larger
scale Pan-STARRS, which will detect every object with a red
magnitude brighter than $24^{th}$ magnitude.  It is estimated that
approximately 10$^5 $ Jovian Trojans exist with red magnitudes
lower than 24, all of which will be detected using Pan-STARRS \citep{jewitt03,durech}.
Our results suggest that Pan-STARRS will reveal between 6000 and 10,000 contact
binary systems among the Trojan clouds.  

\begin{deluxetable}{lccccr}
\tablecolumns{6}
\tablewidth{0pc}
\tabletypesize{\scriptsize}
\tablecaption{Likely Contact Binary Trojans}
\tablehead{
\colhead{Asteroid} 
& \colhead{$\bar m(1,1,0)$\tablenotemark{a}} 
& \colhead{D$_e$ [km]\tablenotemark{b}} 
& \colhead{P [hr]}
& \colhead{$\Delta$m\tablenotemark{c}}
& \colhead{$\rho$ [kg/m$^3$]}}
\startdata
(17365) 	&	10.76	&92		&12.672		&0.98 &780	\\
(29314) 	&	11.80	&32		& 15.035	&1.05 &590	\\
624 Hektor&	7.37	&$350 \times 210$& 6.921	&1.10 &2200	\\
\enddata
\tablenotetext{a}{Mean Absolute Magnitude (see Equation 1)}
\tablenotetext{b}{Effective Diameter (see Equation 2)}
\tablenotetext{c}{Maximum Photometric Range}
\end{deluxetable}\label{table: obs_journal}

\section{Summary}
Sparsely sampled lightcurve measurements were used to
statistically study the photometric variations of 114 Jovian Trojan asteroids.
Objects with large photometric ranges were targeted for follow-up
in this survey, and are considered as candidate contact binary systems.

\begin{enumerate}
\item The sparse sampling technique successfully confirmed known photometric 
ranges of both 944 Hidalgo (0.58 $\pm$ 0.02 magnitudes) and 2674 Pandarus 
(0.50 $\pm$ 0.01 magnitudes).

\item Two of the 114 observed Trojans,  asteroids (17365) and (29314), were found to show
 photometric ranges larger than expected for rotationally deformed equilibrium figures, and were targeted for dense follow-up lightcurve observations.
The resulting ranges (0.98 $\pm$ 0.02 mag
and 1.05 $\pm$ 0.03 mag, respectively) and long rotation periods (12.672 $\pm$ 0.019 hr
and 15.035 $\pm$ 0.007 hr) of these two Trojans are consistent with a contact binary 
structure for each object.

\item Roche binary models give densities of $780^{+50}_{-80}$ kg/m$^3$ for asteroid (17365) and
 $590^{+40}_{-80}$ kg/m$^3$ 
for asteroid (29314), suggestive
of porous interiors.

\item If (17365) and (29314) are indeed contact binaries, then we estimate from our survey
that the binary fraction of the Jovian Trojans is $\sim$6\% to 10\% or more.  The total binary fraction (including
both wide and close pairs) must be higher.

\end{enumerate}

\acknowledgments
We thank John Dvorak, Daniel Birchall, Dave Brennan and Ian Renaud-Kim for 
operating the UH telescope and Henry Hsieh for assisting with the observations 
both in Taiwan and Honolulu.
We are grateful for the assistance and expertise of the Lulin Observatory 
staff, in particular Wen-Ping Chen, Chung-Ming Ko and HC Lin.
Support for this work by a grant to D.J. from NASA's Origins
Program is greatly appreciated.

\clearpage

\end{document}